\documentclass[12pt]{article}
\usepackage{amsmath}
\usepackage{cite}
\usepackage{graphicx}

\csname @addtoreset\endcsname{equation}{section}
\newcommand{\eq}{\begin{equation}}
\newcommand{\feq}{\end{equation}}
\newcommand{\eqn}{\begin{eqnarray}}
\newcommand{\feqn}{\end{eqnarray}}
\newcommand{\arr}{\begin{eqnarray*}}
\newcommand{\farr}{\end{eqnarray*}}

\input{tcilatex}

\begin{document}

\begin{titlepage}
\begin{flushright}
CAMS/02-04\\
hep-th/0207128\\
\end{flushright}
\vspace{.3cm}
\begin{center}
\renewcommand{\thefootnote}{\fnsymbol{footnote}}
{\Large \bf Magnetic Branes 
in $d$-Dimensional AdS Einstein-Maxwell Gravity}
\vskip 2.5 cm
{\large\bf  W.~A.~Sabra\footnote{email: ws00@aub.edu.lb}}
\renewcommand{\thefootnote}{\arabic{footnote}}
\setcounter{footnote}{0}
\vskip 1.5cm
{\small Center for Advanced Mathematical Sciences (CAMS)
and\\
Physics Department, American University of Beirut, Lebanon.\\}
\end{center}
\vfill
\begin{center}
{\bf Abstract}
\end{center}
We present a class of magnetically charged brane solutions for the theory 
of Einstein-Maxwell gravity with a negative cosmological constant in $d$\ $\geq 4$ spacetime dimensions.
\end{titlepage}

\baselineskip=20pt \ \ 

\section{Introduction}

There has been lots of interest in the study of gravitational configurations
in anti-de Sitter spaces. This, to a large extent, has been motivated by the
conjectured equivalence between string theory on anti-de~Sitter (AdS) spaces
(times some compact manifold) and certain superconformal gauge theories
living on the boundary of AdS \cite{ads}. Classical AdS gravity solutions
can furnish important information on the dual gauge theory in the large $N$
limit, $N$ being the rank of the gauge group. Moreover, the AdS/CFT
equivalence makes possible the microscopic analysis of the
Bekenstein-Hawking entropy of asymptotically anti-de~Sitter gravitational
solutions. For example, the central charge of the AdS$_{3}$ asymptotic
symmetry algebra \cite{brown} has been used to count the microstates giving
rise to the BTZ black hole entropy \cite{strominger}. \newline
Given the AdS/CFT correspondence, it becomes important to construct bulk
gravitational solutions to the theory of Einstein gravity with a negative
cosmological constant. Although many brane solutions in ungauged
supergravity theories or gravity theories without a cosmological constant
are known, the same can not be said concerning the corresponding objects in
the AdS cases. Black hole solutions to Einstein's equations with a negative
cosmological constant were constructed a long time ago in \cite{carter}.
Also, black holes with various topologies of the horizon (planar, toroidal
or an arbitrary genus Riemann surface) were discussed in \cite{Lemos, Hu,
Ca, Be, Mann, Vanzo, Brill, rot}. Generalizations to $d$ dimensional anti-de
Sitter black holes with `horizons' given by $(d-2)$-dimensional compact
Einstein spaces were also considered in \cite{Birm}.

In the context of gauged supergravity models, black holes and brane
solutions were discussed in many places (see for example \cite{romans, ck,
cs,london, perry, klemm, bcs2, csb, kkk, chamsabra, ks, duff, dufftasi,
cveduff, gaun, liu2}). It must be noted that Einstein-Maxwell theories with
a negative cosmological constant with spacetime dimensions less than six can
be obtained as a consistent truncation of certain gauged supergravity
models. The main purpose of this work is the study of magnetic brane
solutions with various topologies in anti-de Sitter $d$-dimensional gravity
coupled to a vector potential. We find brane solutions with a magnetic
charge given in terms of the inverse of the cosmological constant. We also
obtain magnetic solutions given in terms of a product of a $(d-2)$%
-dimensional solution for pure Einstein gravity with a negative cosmological
constant and a two dimensional Einstein space.

\section{Branes in $d$-dimensional AdS E-M Gravity}

We start our discussion by considering the anti-de Sitter Einstein-Maxwell
theories in $d$ \ $(\geq 4)$ dimensions. The action for our theory is given
by

\begin{equation}
S=-\frac{1}{16\pi G_{d}}\int_{M}d^{d}x\sqrt{-g}\left( R+(d-1)(d-2)l^{2}-%
\frac{(d-2)}{2(d-3)}F^{2}\right) .  \label{action}
\end{equation}
Here $G_{d}$ is Newton's constant in $d$ dimensions, $R$ is the scalar
curvature and $F^{2}=F_{\mu \nu }F^{\mu \nu },$ with $F_{\mu \nu }$ being
the field strength of an abelian gauge field $A_{\mu }.$ Here we have
expressed the cosmological constant by $\Lambda =-(d-1)(d-2)l^{2}.$

The equations of motion for the metric and the gauge field resulting from
the action (\ref{action}) are given by

\begin{eqnarray}
R_{\mu \nu } &=&\frac{(d-2)}{(d-3)}F_{\mu \rho }F_{\nu }^{\text{ \ }\rho }-%
\frac{1}{2(d-3)}g_{\mu \nu }F^{2}-(d-1)l^{2}g_{\mu \nu },  \label{motioneq}
\\
\partial _{\mu }\left( \sqrt{-g}F^{\mu \nu }\right) &=&0.  \label{gf}
\end{eqnarray}
We consider general $(d-4)$-magnetic brane solutions whose metric can be
written in the following form 
\begin{equation}
ds^{2}=-e^{2V}(-dt^{2}+\sum dz_{\alpha
_{i}}^{2})+e^{2U}dr^{2}+N(r)^{2}d\Omega ^{2},  \label{met}
\end{equation}
where the functions $U$ and $V$ are assumed to depend only on the transverse
coordinate $r$ and we consider either $N(r)=r$ or $N(r)=\mathcal{N}=$
constant. The index $\alpha _{i}$ labels the worldvolume coordinates with $i$
ranging from $1$ to $(d-4).$ Moreover, $d\Omega ^{2}$ denotes the metric of
a two-manifold of a constant Gaussian curvature $k$. In the following, and
without loss of generality, we shall consider the cases $k=0,\pm 1$. Clearly
our two-dimensional space is a quotient space of the universal coverings $%
S^{2}$ ($k=1$), $H^{2}$ ($k=-1$) or $E^{2}$ ($k=0$). Explicitly, we choose 
\begin{equation}
d\Omega ^{2}=d\theta ^{2}+f^{2}d\phi ^{2},
\end{equation}
where

\begin{equation}
f(\theta )=\left\{ 
\begin{array}{l@{\quad,\quad}l}
\sin \theta & k=1, \\ 
1 & k=0, \\ 
\sinh \theta & k=-1.
\end{array}
\right.
\end{equation}

\subsection{Solutions with Spherical, Flat and Hyperbolic Transverse Space}

Using the general form of the Ricci curvature given in the appendix with $%
N(r)=r$, we obtain

\begin{eqnarray}
R_{tt} &=&-R_{z_{\alpha _{i}}z_{\alpha _{i}}}=e^{2V-2U}\left[ (d-3)V^{\prime
2}+V^{\prime \prime }-U^{\prime }V^{\prime }+\frac{2}{r}V^{\prime }\right] ,
\notag \\
R_{rr} &=&-(d-3)(V^{\prime \prime }+V^{\prime 2}-U^{\prime }V^{\prime })+%
\frac{2}{r}U^{\prime },  \notag \\
R_{\theta \theta } &=&\frac{1}{f^{2}}R_{\phi \phi }=e^{-2U}\left[
-(d-3)rV^{\prime }+rU^{\prime }+ke^{2U}-1\right] .  \label{sphericaleq}
\end{eqnarray}
Our brane solutions carry magnetic charge and for the gauge field strength
we take

\begin{equation}
F_{\theta \phi }=\pm kqf,\text{\ }  \label{gaugefield}
\end{equation}
where $q$ is a constant related to magnetic charge. Then using (\ref
{sphericaleq}) and (\ref{gaugefield}) in (\ref{motioneq}), we obtain the
following differential equations 
\begin{eqnarray}
(d-3)V^{\prime 2}+V^{\prime \prime }-U^{\prime }V^{\prime }+\frac{2V^{\prime
}}{r} &=&e^{2U}\left[ \frac{k^{2}q^{2}}{(d-3)r^{4}}+(d-1)l^{2}\right] ,
\label{grone} \\
-(d-3)(V^{\prime \prime }+V^{\prime 2}-U^{\prime }V^{\prime })+\frac{%
2U^{\prime }}{r} &=&-e^{2U}\left[ \frac{k^{2}q^{2}}{(d-3)r^{4}}+(d-1)l^{2}%
\right] ,  \label{grtwo} \\
-(d-3)rV^{\prime }+rU^{\prime }+ke^{2U}-1 &=&e^{2U}\left[ \frac{k^{2}q^{2}}{%
r^{2}}-r^{2}(d-1)l^{2}\right] .  \label{grthree}
\end{eqnarray}
To proceed in solving the resulting equations of motion we first write

\begin{equation}
V^{\prime }=e^{U}W.  \label{an}
\end{equation}
Then by adding up the equations (\ref{grone}) and (\ref{grtwo}), we obtain
the following differential equation 
\begin{equation}
\frac{2}{r}W+(4-d)W^{\prime }+\frac{2}{r}U^{\prime }e^{-U}=0.
\label{sumequation}
\end{equation}
Equation (\ref{grthree}) provides an expression for $W$ in terms of $U$
which reads

\begin{equation}
W=\frac{e^{U}}{(d-3)}\left[ -\frac{k^{2}q^{2}}{r^{3}}+r(d-1)l^{2}+\frac{1}{r}%
(k-e^{-2U})+U^{\prime }e^{-2U}\right] .  \label{fequation}
\end{equation}
Upon substituting (\ref{fequation}) in (\ref{sumequation}), we finally
obtain a differential equation for $U,$

\begin{eqnarray}
&&\frac{k^{2}q^{2}\left( 10-3d\right) }{r^{4}}+\left( 6-d\right)
(d-1)l^{2}+\left( 4-d\right) U^{\prime }\left[ -\frac{k^{2}q^{2}}{r^{3}}%
+r(d-1)l^{2}+\frac{k}{r}\right]  \notag \\
&&+e^{-2U}\left[ \left( 4-d\right) \left( U^{\prime \prime }-U^{\prime
2}\right) +\frac{dU^{\prime }}{r}+\frac{(2-d)}{r^{2}}\right] +\frac{k(d-2)}{%
r^{2}}  \notag \\
&=&0.  \label{nice}
\end{eqnarray}
The differential equation (\ref{nice}) admits a simple solution when the
magnetic charge is expressed in terms of the cosmological constant by $q=%
\frac{1}{(d-2)l}.$ In this case our solution is given by 
\begin{equation}
e^{-U}=lr+\frac{kq}{r}=lr+\frac{k}{(d-2)lr}.  \label{u}
\end{equation}
The function $W$ can be easily determined from equation (\ref{fequation})
and we get 
\begin{equation}
W=\left[ l-\frac{k}{(d-3)(d-2)lr^{2}}\right] .  \label{f}
\end{equation}
Using (\ref{an}), (\ref{u}) and (\ref{f}) we can finally solve for $V$ and
get 
\begin{equation}
e^{V}=lr\left[ 1+\frac{k}{(d-2)l^{2}r^{2}}\right] ^{\frac{1}{2}\left( \frac{%
d-2}{d-3}\right) }.
\end{equation}
Therefore, $(d-4)$-magnetic branes with spherical transverse space are
described by the metric and gauge field 
\begin{eqnarray}
ds^{2} &=&(lr)^{2}\left[ 1+\frac{1}{(d-2)l^{2}r^{2}}\right] ^{\frac{d-2}{d-3}%
}\left( -dt^{2}+\sum dz_{\alpha _{i}}^{2}\right) +\left[ lr+\frac{1}{(d-2)lr}%
\right] ^{-2}dr^{2}  \notag \\
&&+r^{2}\left( d\theta ^{2}+\sin ^{2}\theta d\phi ^{2}\right) ,  \notag \\
F_{\theta \phi } &=&\pm \frac{\sin \theta }{\text{\ }(d-2)l}.  \label{sm}
\end{eqnarray}
Note that for the $d=5$ spherical case we obtain the magnetic string which
has been considered in \cite{chamsabra}. Also for $d=4$ (in which case there
are no $z_{\alpha _{i}}$ coordinates), we obtain the ``cosmic monopole''
solution which was first discovered by Romans\cite{romans}. The cases of
four and five dimensions can be embedded in $N=2$ gauged supergravity and
the corresponding magnetic solutions were shown to preserve some
supersymmetry \cite{romans, cs, chamsabra}. For magnetic branes with
hyperbolic transverse space (i. e., $f=\sinh \theta )$ we have \footnote{%
It must be noted that the hyperbolic magnetic brane metric can be obtained
from (\ref{sm}) by making the following substitutions :
\par
$t\rightarrow it,$ \ \ \ \ \ $r\rightarrow ir,$ \ \ \ \ $z\rightarrow iz,$ \
\ \ \ \ \ $\theta \rightarrow i\theta ,$ $\ \ \ \ \ \ \ \phi \rightarrow
\phi $%
\par
{}} 
\begin{eqnarray}
ds^{2} &=&(lr)^{2}\left[ 1-\frac{1}{(d-2)l^{2}r^{2}}\right] ^{\frac{d-2}{d-3}%
}\left( -dt^{2}+\sum dz_{\alpha _{i}}^{2}\right) +\left[ lr-\frac{1}{(d-2)lr}%
\right] ^{-2}dr^{2}  \notag \\
&&+r^{2}\left( d\theta ^{2}+\sinh ^{2}\theta d\phi ^{2}\right) .
\end{eqnarray}
These solutions are the generalizations to $d$ dimensions of the hyperbolic
five-dimensional string solution constructed in \cite{ks}. We note that,
unlike the spherical magnetic brane which contains a naked singularity, the
hyperbolic black magnetic brane has an event horizon at $r=1/(l\sqrt{d-2})$.
This is analogous to the AdS$_{4}$ and AdS$_{5\text{ }}$cases discussed in 
\cite{ck, cs, chamsabra, ks}.

The solution with flat transverse space ($f=1),$ which is locally AdS$_{d},$
is a limiting case of a family of black branes, whose metric is given by 
\begin{equation}
ds^{2}=-e^{2V}dt^{2}+e^{-2V}dr^{2}+l^{2}r^{2}\sum dz_{\alpha
_{i}}^{2}+r^{2}(d\theta ^{2}+d\phi ^{2}),  \label{birmin}
\end{equation}
where 
\begin{equation*}
e^{2V}=-\frac{m}{r^{d-3}}+l^{2}r^{2}
\end{equation*}
and $m$ is a constant related to the mass of the magnetic brane. If $%
lz_{\alpha _{i}}$ is considered as a coordinate of the transverse space,
then the solutions (\ref{birmin}) can also be interpreted as black hole
solutions which were actually constructed in \cite{Birm}.

\subsection{Constant Warp Factor}

Here we consider the case where the warp factor $N(r)$ $=\mathcal{N=}$
constant and we take $F_{\theta \phi }=\pm qf$. Inspecting the equations of
motion (\ref{motioneq}) in this case, we find that our solution factors into
a product of two spaces $M_{d-2}$ and $M_{2}$ with metrics which we denote
by $g_{mn}$ and $g_{\alpha \beta }$ respectively. The curvature tensors of $%
M_{d-2}$ and $M_{2}$ are respectively given by \ \ \ \ 
\begin{eqnarray}
R_{mn} &=&-g_{mn}\left[ \frac{q^{2}}{(d-3)\mathcal{N}^{4}}+(d-1)l^{2}\right]
=-c_{1}g_{mn},\text{ \ \ \ } \\
R_{\alpha \beta } &=&g_{\alpha \beta }\left[ \frac{q^{2}}{\mathcal{N}^{4}}%
-(d-1)l^{2}\right] .
\end{eqnarray}
This implies that $M_{d-2}$ is given by a $(d-2)$-dimensional Einstein space
with a negative cosmological constant. For various topologies of $%
M_{2},(k=0,\pm 1),$ the magnetic charge is given by 
\begin{equation}
q^{2}=(d-1)l^{2}\mathcal{N}^{4}+k\mathcal{N}^{2},
\end{equation}
and the Ricci curvature of $M_{d-2}$ is thus given by 
\begin{equation}
R_{mn}=-g_{mn}\left[ \frac{(d-1)(d-2)}{\left( d-3\right) }l^{2}+\frac{k}{%
\left( d-3\right) \mathcal{N}^{2}}\right] .
\end{equation}
If we restrict ourselves to the metric ansatz (\ref{met}) with constant warp
factor, then the curvature tensor of $M_{d-2}$ is given by

\begin{eqnarray}
R_{tt} &=&-R_{z_{\alpha _{i}}z_{\alpha _{i}}}=e^{2V-2U}\text{ }\left[
(d-3)V^{\prime 2}-U^{\prime }V^{\prime }+V^{\prime \prime }\right] ,  \notag
\\
R_{rr} &=&-\text{ }(d-3)\left( V^{\prime 2}+U^{\prime }V^{\prime }-V^{\prime
\prime }\right) .
\end{eqnarray}
The equations of motion in this case are 
\begin{eqnarray}
\text{ }(d-3)V^{\prime 2}-U^{\prime }V^{\prime }+V^{\prime \prime }
&=&c_{1}e^{2U},  \notag \\
(d-3)\left( V^{\prime 2}+U^{\prime }V^{\prime }-V^{\prime \prime }\right)
&=&c_{1}e^{2U}.  \label{fa}
\end{eqnarray}
As an example, we take $q=\frac{1}{(d-2)l}$ and $f=\sinh \theta $. This
implies that

\begin{equation}
\mathcal{N}\text{ }^{2}=\frac{1}{(d-2)l^{2}}.
\end{equation}
For this value of $\mathcal{N},$ \ one gets for the curvature of $M_{d-2}$ \
\ \ \ 
\begin{equation}
R_{mn}=-g_{mn}\frac{(d-2)^{2}l^{2}}{\left( d-3\right) },
\end{equation}
and the equations of motion (\ref{fa}) can be seen to be satisfied for the
choice

\begin{equation}
V^{\prime }=l\left( \frac{d-2}{d-3}\right) e^{U}\text{.}  \label{hor}
\end{equation}
The solution thus obtained describes the near horizon geometry of our
hyperbolic magnetic brane solution which is given by AdS$_{(d-2)}\times $ $%
H^{2}.$

As examples of $M_{d-2},$ which are solutions of Einstein's equations with a
negative cosmological constant in $(d-2)$ dimensions, we can take the
solutions presented in \cite{Birm}. For a $d^{\prime }$-dimensional space,
those spacetimes are given by 
\begin{equation}
ds^{2}=-\left( k^{\prime }-\frac{m}{r^{d^{\prime }-3}}+\lambda
^{2}r^{2}\right) dt^{2}+\left( k^{\prime }-\frac{m}{r^{d^{\prime }-3}}%
+\lambda ^{2}r^{2}\right) dr^{2}+r^{2}h_{ij}(x)dx^{i}dx^{j}.
\end{equation}
The metric function $h_{ij}$ is a function of the coordinates $x^{i}$ $%
(i=1,...,(d^{\prime }-2))$ and is referred to as the horizon metric. In \cite
{Birm}, it was demonstrated that the above spacetimes are Einstein spaces
with a negative cosmological constant, namely 
\begin{equation}
R_{\mu \nu }=-(d^{\prime }-1)\lambda ^{2}g_{\mu \nu },  \label{j}
\end{equation}
provided that the horizon metric is an Einstein space 
\begin{equation}
R_{ij}(h)=(d^{\prime }-3)k^{\prime }\;h_{ij},\text{ }k^{\prime }=0,\pm 1.
\label{ji}
\end{equation}

\section{Summary}

To sum up, we presented new magnetic $(d-4)$-brane solutions for the theory
of $d$-dimensional Einstein-Maxwell theory with a negative cosmological
constant. We discussed the cases of magnetically charged brane solutions for
spherical, hyperbolic and flat transverse space. We found that the magnetic
charge satisfies a ``quantization relation'' in terms of \ the inverse of
the square of the cosmological constant. The solutions found are solitonic
objects in the sense that the flat-space limit $l\rightarrow 0$ does not
exist. Our solutions can be summarized by

\begin{eqnarray}
ds^{2} &=&(lr)^{2}\left[ 1+\frac{k}{(d-2)l^{2}r^{2}}\right] ^{\frac{d-2}{d-3}%
}\left( -dt^{2}+\sum dz_{\alpha _{i}}^{2}\right) +\frac{1}{(lr)^{2}\left[ 1+%
\frac{k}{(d-2)l^{2}r^{2}}\right] ^{2}}dr^{2}  \notag \\
&&+r^{2}\left( d\theta ^{2}+f^{2}d\phi ^{2}\right) .\text{\ \ \ \ \ \ \ \ }
\label{gen}
\end{eqnarray}
with the gauge field strength given by $F_{\theta \phi }=\pm \frac{kf}{%
\left( d-2\right) l}.$ We note that it would be interesting to study more
general solutions for which (\ref{gen}) are special limiting cases. We also
described the cases where we have a constant warp factor and there we found
that our solution factors into a product of two spaces, one of which is
given by a $(d-2)$-dimensional solution of the theory of Einstein gravity
with a negative cosmological constant. The other factor of the product space
is a two-dimensional Einstein manifold.\vfill\eject

\section{Appendix}

Our conventions are as follows. General curved indices are labelled by $\mu
, $ $\nu ,....$, time and transverse coordinates are labelled by $(t,$ $r,$ $%
\theta ,$ $\phi )$ with the corresponding flat indices labelled by $(0,$ $1,$
$2,$ $3).$ Worldvolume coordinates are labelled by $z_{\alpha _{i}}$, (with $%
i=1,\cdots ,d-4)$ with corresponding flat indices given by $\ \alpha _{i}.$
Consider the following metric

\begin{equation}
ds^{2}=e^{2V(r)}(-dt^{2}+\sum dz_{\alpha
_{i}}^{2})+e^{2U(r)}dr^{2}+N(r)^{2}\left( d\theta ^{2}+f^{2}d\phi
^{2}\right) .
\end{equation}
The dual basis of one-forms is given by

\begin{equation}
\vartheta ^{0}=e^{V}dt,\text{ \ \ \ }\vartheta ^{\alpha
_{i}}=e^{V}dz^{\alpha _{i}},\text{ \ \ \ }\vartheta ^{1}=e^{U}dr\text{,\ \ \
\ \ }\vartheta ^{2}=Nd\theta ,\text{ \ \ \ }\vartheta ^{3}=Nf\text{\ }d\phi .
\end{equation}
Exterior differentiation of the above forms gives

\begin{eqnarray}
d\vartheta ^{0} &=&V^{\prime }e^{-U}\text{ }\vartheta ^{1}\wedge \vartheta
^{0},  \notag \\
d\vartheta ^{\alpha _{i}} &=&V^{\prime }e^{-U}\vartheta ^{1}\wedge \vartheta
^{\alpha _{i}},  \notag \\
d\vartheta ^{1} &=&0,  \notag \\
d\vartheta ^{2} &=&\frac{N^{\prime }}{N}e^{-U}\text{ }\vartheta ^{1}\wedge
\vartheta ^{2},  \notag \\
d\vartheta ^{3} &=&\frac{N^{\prime }}{N}e^{-U}\text{ }\vartheta ^{1}\wedge
\vartheta ^{3}+\frac{\partial _{\theta }f\text{ }}{Nf}\vartheta ^{2}\wedge
\vartheta ^{3}.
\end{eqnarray}
Here the prime sign denotes differentiation with respect to $r.$ Using

\begin{equation}
d\vartheta ^{a}+w_{\text{ }b}^{a}\wedge \vartheta ^{b}=0,
\end{equation}
we obtain for the connection forms the following

\begin{eqnarray}
w_{\text{ }1}^{0} &=&V^{\prime }e^{-U}\text{ }\vartheta ^{0},\text{ \ \ }w_{%
\text{ }1}^{\alpha _{i}}=V^{\prime }e^{-U}\text{ }\vartheta ^{\alpha _{i}},%
\text{ \ \ }w_{\text{ }1}^{2}=\frac{N^{\prime }}{N}e^{-U}\text{ }\vartheta
^{2},\text{ \ \ }  \notag \\
w_{\text{ }1}^{3} &=&\frac{N^{\prime }}{N}e^{-U}\text{ }\vartheta ^{3},\text{
\ \ }w_{\text{ }2}^{3}=\frac{1}{N}\frac{\partial _{\theta }f\text{ }}{f}%
\text{ }\vartheta ^{3}.
\end{eqnarray}
Upon using 
\begin{equation*}
dw_{\text{ }b}^{a}+w_{\text{ }e}^{a}\wedge w_{\text{ }b}^{e}=\frac{1}{2}R_{%
\text{ }bef}^{a}\vartheta ^{e}\wedge \vartheta ^{f},
\end{equation*}
we obtain the components of the Ricci tensor $R_{ab}=g^{ef}R_{eafb}$ 
\begin{eqnarray}
R_{00} &=&-R_{\alpha _{i}\alpha _{i}}=e^{-2U}\left[ \text{ }(d-3)V^{\prime
2}-U^{\prime }V^{\prime }+V^{\prime \prime }+\frac{2V^{\prime }N^{\prime }}{N%
}\right] ,  \notag \\
R_{11} &=&-e^{-2U}\text{ }\left[ (d-3)\left( V^{\prime 2}-U^{\prime
}V^{\prime }+V^{\prime \prime }\right) -\frac{2}{N}(U^{\prime }N^{\prime
}-N^{\prime \prime })\right] ,  \notag \\
R_{22} &=&R_{33}=\frac{1}{N}e^{-2U}\text{ }\left[ -(d-3)V^{\prime }N^{\prime
}+(U^{\prime }N^{\prime }-N^{\prime \prime })+\text{ }\frac{1}{N}%
(ke^{2U}-N^{\prime 2})\right] .
\end{eqnarray}

\end{document}